# Photonic Band Gap in 1D Multilayers Made by Alternating $SiO_2$ or PMMA with $MoS_2$ or $WS_2$ monolayers

Diana Gisell Figueroa del Valle[1,2], Eduardo Aluicio-Sarduy[1,3], Francesco Scotognella[1,2,4]*

[1] *Center for Nano Science and Technology@PoliMi, Istituto Italiano di Tecnologia, Via Giovanni Pascoli, 70/3, 20133, Milan, Italy*
[2] *Dipartimento di Fisica, Politecnico di Milano, Piazza Leonardo da Vinci 32, 20133 Milano, Italy*
[3] *Dipartimento di Chimica, Materiali e Ingegneria Chimica "Giulio Natta", Politecnico di Milano, Piazza Leonardo da Vinci 32, 20133 Milano, Italy*
[4] *Istituto di Fotonica e Nanotecnologie CNR, Piazza Leonardo da Vinci 32, 20133 Milano*
* *Corresponding author at: Dipartimento di Fisica, Politecnico di Milano, Piazza Leonardo da Vinci 32, 20133 Milano, Italy. E-mail address: francesco.scotognella@polimi.it (F. Scotognella).*

**Abstract**
Atomically thin molybdenum disulphide ($MoS_2$) and tungsten disulphide ($WS_2$) are very interesting two dimensional materials for optics and electronics. In this work we show the possibility to obtain one-dimensional photonic crystals consisting of low-cost and easy processable materials, as silicon dioxide ($SiO_2$) or poly methyl methacrylate (PMMA), and of $MoS_2$ or $WS_2$ monolayers. We have simulated the transmission spectra of the photonic crystals using the transfer matrix method and employing the wavelength dependent refractive indexes of the materials. This study envisages the experimental fabrication of these new types of photonic crystals for photonic and light emission applications.

**Keywords**: photonic crystals; transfer matrix method; transition metal dichalcogenides; $MoS_2$ monolayer; $WS_2$ monolayer.

**Introduction**
The research on atomically thin molybdenum disulphide ($MoS_2$) and tungsten disulphide ($WS_2$) is gaining considerable attention, since they are promising for several applications in optics and electronics [1–3]. In literature there are several reports on the optical properties, time resolved photogeneration, and luminescence of few-layers of $MoS_2$ [4], $MoS_2$ monolayer [5–7], and $WS_2$ monolayer [8]. Very recently, the development of population inversion, very useful for laser applications, has been observed in $WS_2$ [9].
Photonic crystals are composite materials in which two compounds, with a different refractive index, are alternated and the arising periodicity is comparable with the wavelength of light. Such periodicity gives rise to energy regions where the photons are not allowed to propagate in the materials, called photonic band gap [10–12]. Multilayered structures are a particular case of one-dimensional photonic crystals, which are easy to fabricate with several methods [13–15], and can be employed for different applications, as sensors [16], electro-optic switches [17], and lasers [18]. In literature there are interesting reports on one-dimensional photonic crystals containing $MoS_2$ [19] and on microcavities in which a $MoS_2$ monolayer is embedded [20,21]. However, a study of the transmission properties of one-dimensional multilayer photonic crystals, where atomically thin layers of different semiconducting transition-metal dichalcogenides are employed in the periodic structures, is still missing.
Here we describe the realization of a one-dimensional photonic crystal that is made with materials that are usually employed for these types of crystals, as silicon dioxide ($SiO_2$) or poly methyl methacrylate (PMMA), and transition metal dichalcogenides as $MoS_2$ or $WS_2$. The photonic crystals consist of multilayers and we have simulated the transmission spectra of the photonic crystals using the transfer matrix method and utilizing the wavelength dependent refractive indexes of the materials. We have found that a photonic band gap arises with atomically thin layers of MoS$_2$ or WS$_2$.

**Methods**
We have considered an air/multilayer/glass system. The wavelength dependent refractive index of SiO$_2$ is taken from Ref. [22], and the corresponding Sellmeier equation is:
$$n_{SiO_2}(\lambda) = \left(1 + \frac{0.6962\lambda^2}{\lambda^2-0.0684^2} + \frac{0.4080\lambda^2}{\lambda^2-0.1162^2} + \frac{0.8975\lambda^2}{\lambda^2-9.8962^2}\right)^{1/2} \quad (1).$$
Instead, the refractive index of PMMA is taken from Ref. [23] and the Sellmeier equation is:
$$n_{PMMA}(\lambda) = \left(1 + \frac{0.99654\lambda^2}{\lambda^2-0.00787} + \frac{0.18964\lambda^2}{\lambda^2-0.02191} + \frac{0.00411\lambda^2}{\lambda^2-3.85727}\right)^{1/2} \quad (2)$$
The unit for the wavelength in the two Sellmeier equations is micrometer.

The wavelength dependent dielectric function (real and imaginary parts) of MoS$_2$ is taken from Ref. [24], while the one of WS$_2$ from Ref. [25], considering the refractive index of MoS$_2$/WS$_2$ $n_{MoS_2/WS_2} = \sqrt{\varepsilon_{MoS_2/WS_2}}$.

The simulation of the transmission spectra of the multilayers have been performed by using the transfer matrix method, exhaustively described in Refs. [26–28].

**Results and Discussion**
We have simulated the transmission spectra of multilayer photonic crystals in which the layers of SiO$_2$, or PMMA, are alternated with monolayers of MoS$_2$, or WS$_2$. In Figure 1 we have depicted a sketch of the photonic crystal.

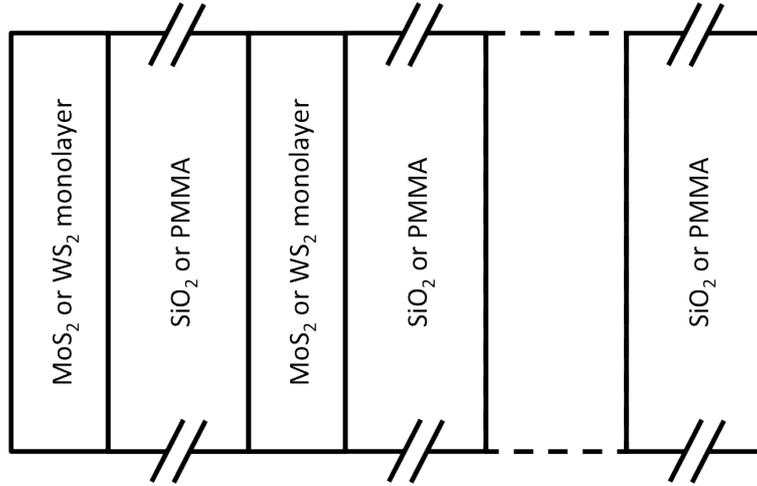

Figure 1. Schemes of the photonic crystals made by alternating layers of SiO$_2$ or PMMA with atomically thin layers of MoS$_2$ or WS$_2$.

We have carefully taken into account the wavelength dependent refractive index of all the employed materials, in order to give an accurate prediction of the transmission spectra of the photonic crystals. For the MoS$_2$/SiO$_2$ multilayer, we have chosen a thickness of the SiO$_2$ layers of 250 nm, while the thickness of the monolayer of MoS$_2$ is 0.65 nm [24]. In Figure 2 we show the transmission spectrum of a MoS$_2$/SiO$_2$ photonic crystal made by 15 bilayers.

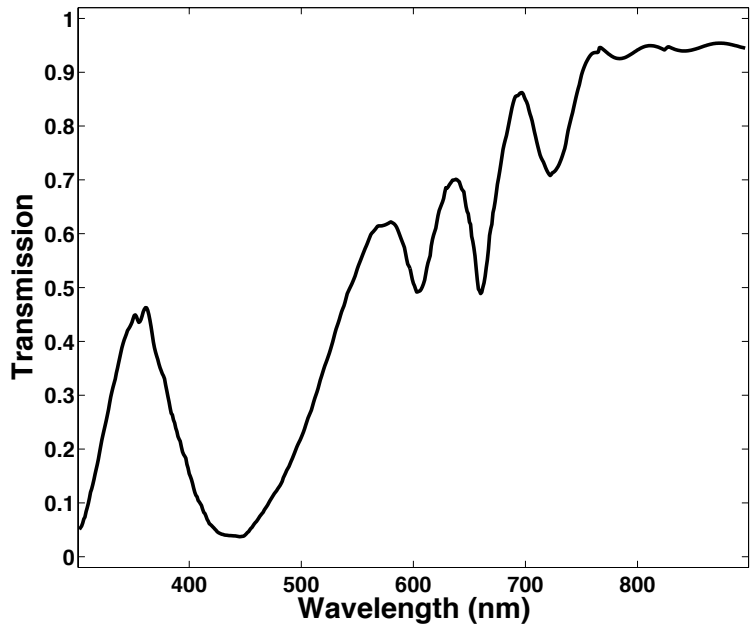

Figure 2. Transmission spectrum of a one-dimensional photonic crystal made alternating 15 layers of $SiO_2$ and 15 monolayers of $MoS_2$.

In the spectrum the transmission valleys at around 450, 605, and 660 nm are related to the absorption of $MoS_2$ (due to the excitonic resonances [29–31]). The valley at about 720 nm is related to a photonic band gap arising from the periodicity of the refractive index.

For the $WS_2/SiO_2$ multilayer the thickness of the $SiO_2$ layers is again 250 nm, and the thickness of the monolayer of $WS_2$ is 0.65 nm [25]. In Figure 3 we show the transmission spectrum of a $WS_2/SiO_2$ photonic crystal made by 15 bilayers. In this multilayer the transmission valleys related to the excitonic resonances of $WS_2$ [8,29] are at 435 nm, 515 nm, and the very intense one at 618 nm. The photonic band gap is observed at about 720 nm.

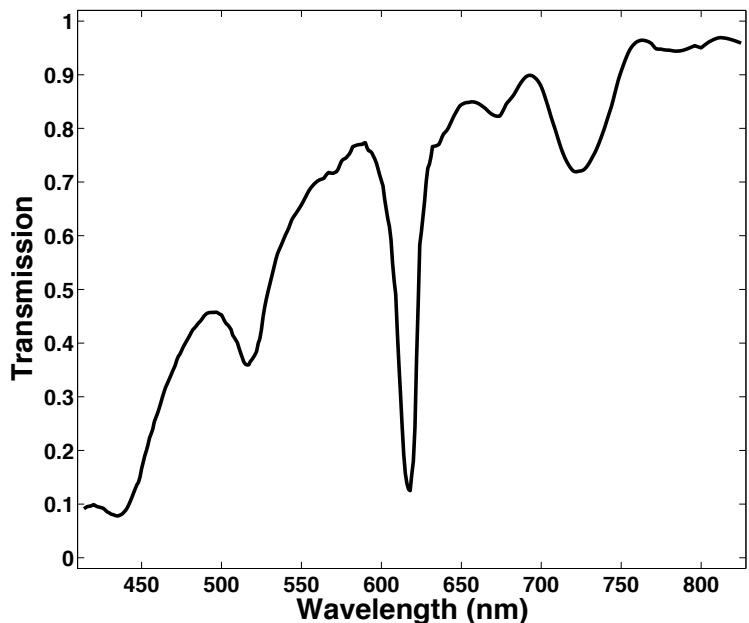

Figure 3. Transmission spectrum of a one-dimensional photonic crystal made alternating 15 layers of $SiO_2$ and 15 monolayers of $WS_2$.

We have also simulated the transmission spectra of photonic crystals in which the transition metal dichalcogenide monolayers are alternated with a plastic material as PMMA. For the $MoS_2$/PMMA multilayer the thickness of the PMMA layers, as for $SiO_2$, is 250 nm, and its transmission spectrum is shown in Figure 4.

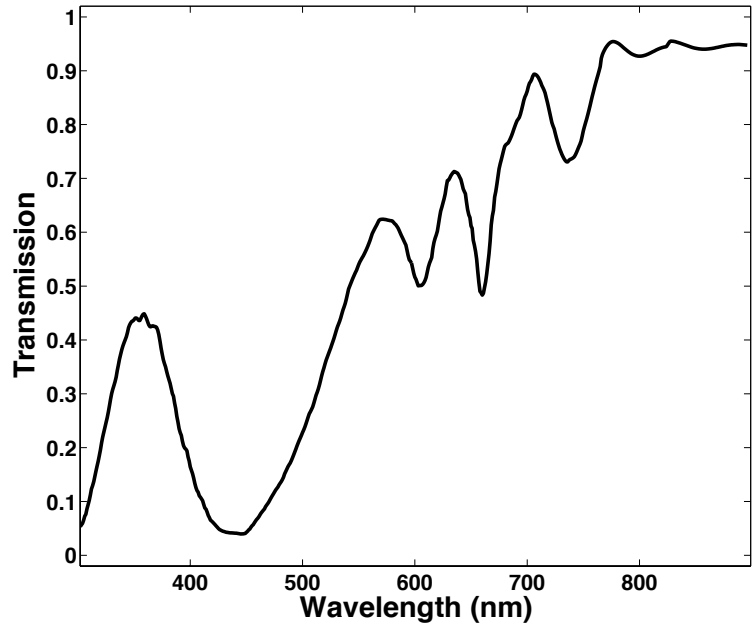

Figure 4. Transmission spectrum of a one-dimensional photonic crystal made alternating 15 layers of PMMA and 15 monolayers of $MoS_2$.

Also in the case of the atomically thin $WS_2$/PMMA multilayer the thickness of the PMMA layers is 250 nm, and its transmission spectrum is shown in Figure 5 (1ML, i.e. a monolayer).

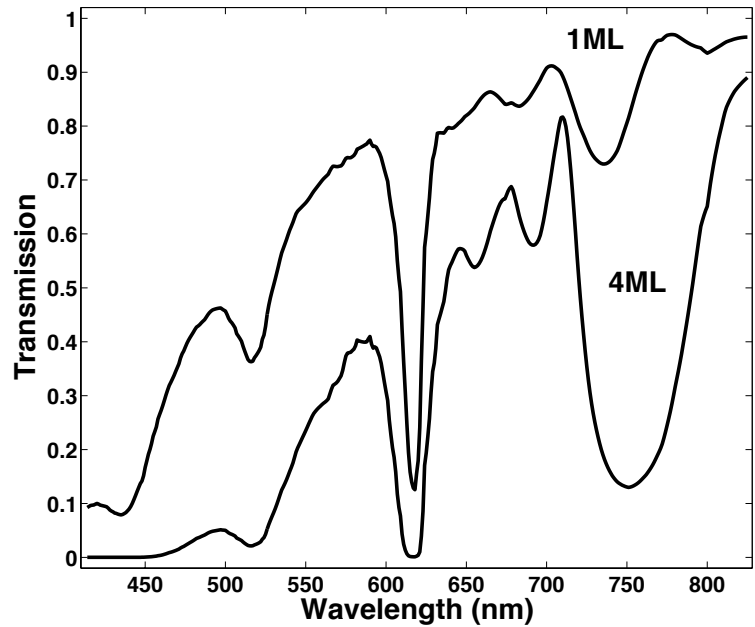

Figure 5. Transmission spectrum of a one-dimensional photonic crystal made alternating 15 layers of PMMA and 15 monolayers of $WS_2$ (1ML) or 15 tetralayers of $WS_2$ (4ML).

It is worth noting the photonic band gaps on the multilayers that we have shown are weak in intensity, with a transmission minimum between 70% and 80%. This is due to the fact the thickness of $MoS_2$ or $WS_2$ monolayers is inherently very small (i.e. 0.65 nm). To obtain a more intense photonic band gap a higher number of bilayers is needed. Another possible solution is to increase the thickness of $MoS_2$ or $WS_2$ in the multilayer, i.e. from a monolayer to a tetralayer. For example, we have also simulated the transmission spectrum of a multilayer photonic crystal where the $WS_2$ consists of a tetralayer, with a thickness of 2.6 nm. In this case, because of the larger thickness of $WS_2$, we have decreased the thickness of PMMA to 245 nm. The transmission spectrum of the photonic crystal is shown in Figure 5 (4ML). We have observed that the photonic band gap is already very strong with 15 bilayers, reaching a transmission, at about 750 nm, of around 0.1.

From an experimental point of view, the photonic crystals can be fabricated by employing different techniques. The deposition of the $SiO_2$ layers can be done by radiofrequency sputtering [14] or chemical vapor deposition [20], while the deposition of the PMMA layers by spin coating [32]. Transition metal dichalcogenide monolayers can be synthetized by ambient pressure chemical vapour deposition and transfer, on the other layers of the photonic crystal, via an aqueous solution transfer method [20,33].

## Conclusion

In summary, we have shown the possibility to realize a one dimensional multilayer photonic crystal consisting of the usual low cost materials for multilayers, as silicon dioxide or PMMA, and atomically thin layers of two different transition metal dichalcogenides, as $MoS_2$ and $WS_2$. We have simulated the transmission spectra of the photonic crystals using the transfer matrix method and taking into account the wavelength dependent refractive indexes of the materials. The study here described envisages the experimental fabrication of these photonic crystals for photonic and light emission applications. For example, the photonic band gap can be tuned in order modulate, and enhance, the luminescence of $MoS_2$ or $WS_2$ monolayers [2,6,7].

## Acknowledgements

D.G.F.d.V. and F.S. want to acknowledge the ITN project 316633 ''POCAONTAS''. F. S. also wants to acknowledges financial support from the Italian Ministry of University and Research (project PRIN 2010-2011 "DSSCX" No. 20104XET32).